\documentstyle[preprint,aps,tighten,epsfig,axodraw,epic,floats,psfig,amssymb]{revtex}
\begin{document}
\newcommand{\onhalf}{{\frac{1}{2}}}
\newcommand{\dpdy}{{\frac{\partial\phi}{\partial y}}}
\newcommand{\dpdz}{{\frac{\partial\phi}{\partial z}}}
\newcommand{\delndx}{{\delta(\vec{n}_i \cdot \vec{w})}}
\newcommand{\theldx}{{\theta(\vec{l}_i \cdot \vec{w})}}
\newcommand{\Omegain}{{\Omega}}
\newcommand{\Omegainn}{{\Omega^2}}
\newcommand{\sphii}{{\sin\alpha_i}}
\newcommand{\cphii}{{\cos\alpha_i}}
\newcommand{\deltwo}{{\delta(z\cos\alpha-y\sin\alpha)}}
\newcommand{\delz}{{\delta(z)}}
\tighten
\preprint{ \vbox{
\hbox{MADPH--00-1196}
\hbox{UPR-906-T}
\hbox{hep-th/0109073}}}
\draft
\title{Self-Tuning and de Sitter Brane Intersections in the 
6-Dimensional Brane Models}
\author{Jing Jiang$^1$\footnote{Current Address: Division of High Energy
Physics, Argonne National Laboratory, Argonne, IL 60439.}
and Tianjun Li$^2$}
\vskip 0.3in
\address{$^1$Department of Physics, University of Wisconsin--Madison,
 WI 53706}
\address{$^2$Department of Physics and Astronomy, \\ 
University of Pennsylvania, Philadelphia, PA 19104}
\vskip 0.15in
\maketitle

\begin{abstract}
{\rm 
We study the self-tuning of general brane junctions and brane
networks on the 6-dimensional space-time. For the general brane junctions,
 there may exist one fine-tuning among the brane tensions. 
For the brane networks,
similar to the 5-dimensional self-tuning brane models, the brane
tensions can be set arbitrarily and 
there exists the singularity for the metric and bulk scalar. And if we
want to
regularize the singularity, we will introduce the fine-tuning 
among the brane tensions. In addition, because
the 4-dimensional cosmological
constant we observe may be positive and very small,
we discuss the brane network
with de Sitter brane intersections by introducing a bulk scalar.
}
\end{abstract}

\section{Introduction}
 There are some unattractive features in the Standard Model (SM) which
may imply the new physics, although the SM is 
 very succesful from the experiments at
LEP and Tevatron.
One of these problems is that the gauge interactions and gravitational 
interaction are not unified. Another is
the gauge hierarchy problem. As we know, several solutions to the
gauge hierarchy problem have
been proposed: the technicolor and compositeness
which lacks calculability; the weak-scale supersymmetry
which is the leading candidate for the extension of the Standard Model
several years ago; and the conformality whose spirit is similar
to that of supersymmetry: one just replaces one symmetry (supersymmetry)
with another (conformal symmetry) above the TeV scale, and both approaches 
predict new physics at TeV scale~\cite{Frampton}.

About three years ago, it was suggested that the large 
compactified extra dimensions may also be a solution to the
gauge hierarchy problem~\cite{AADD}, because a low ($4+n$)-dimensional
Planck scale ($M_X$) may result in the large 4-dimensional Planck scale 
($M_{pl}$) due
to the large physical volume ($V_p^n$)
 of extra dimensions: $M_{pl}^2 = M_X^{2+n} V_p^n$. In addition,
 Randall and Sundrum~\cite{LRRS} proposed another scenario
that the extra dimension is an orbifold,
and  the size of extra dimension is not large
but the 4-dimensional mass scale in the Standard Model is
suppressed by an exponential factor from the 5-dimensional mass
scale due to the exponential warp factor. Furthermore,  they suggested
that
the fifth dimension might be coordinate non-compact~\cite{LRRSN}, 
and there may exist only one
brane with positive tension at origin, however, there  exists the  gauge
hierarchy problem.  The remarkable aspect of
the second scenario is that it gives rise to a localized graviton field.
After that, a lot of 5-dimensional models with 3-branes were built [5-6],
and the models with co-dimension one brane(s) were constructed on 
the 6-dimensional and higher dimensional space-time [7-11].  
  
In above model buildings,  all the models with warp factor in the
metric have negative bulk
 cosmological constant. However, in string theory, it is natural to
take the bulk cosmological constant to be zero since the tree-level 
vacuum energy in the generic critical closed string compactifications
(supersymmetric or not) vanishes. And the zero bulk cosmological
constant is natural in the scenario in which the bulk is supersymmetric
(though the brane need not be), or the quantum corrections to the
bulk are small enough to be neglect in a controlled expansion.
So, how to construct
the models with zero bulk cosmological constant is an interesting
question in the model buildings, because such kinds of models are still 
interesting if the bulk corrections to 
bulk cosmological constant $\Lambda$ were very small, which
can be happened for instance if the supersymmetry breaking is localized in
a small neighborhood of the branes, or if the supersymmetry breaking scale
in the bulk is small enough. Moreover, if all the gauge fields and matter
fields
were confined to the branes, the quantum corrections of these fields to
the
brane tensions might not affect the models with $\Lambda=0$.

One scenario is that we introduce not
only the space-like extra dimension, but also the time-like extra 
dimension~\cite{JUN}.
The good aspect of this approach is that there is no singularity,
however, there exists fine-tuning and might have the
 problems arising from the time-like extra dimension:
unitarity and causality.
The other scenario was proposed where a 
scalar $\phi$, which does not have bulk potential, 
is introduced~\cite{AHDKS}.
In the second scenario, $\phi$ becomes singular at a finite distance along
the extra dimension and the warp factor in the metric vanishes at
singularity.
The good aspect of this approach is that, the brane tension can be set
arbitrary. However, the $Z_2$ symmetric and 4-dimensional Poincare
invariant
solution is unstable under the bulk quantum corrections, and any procedure
which regularizes the singurality will introduce the fine-tuning which the
self-tuning is supposed to avoid~\cite{FLLN}. 
brane~\cite{PBJMCG}. 
Furthermore, a simple no-go theorem~\cite{nogo} relating to the
self-tuning 
solutions to the 
cosmological constant for observers on the brane, which rely on a
singularity 
in an extra dimension, shows that it is impossible to shield the
singularity from the 
brane by a horizon~\cite{CEG}, unless the positive energy condition is
violated 
in the bulk or on the
brane, or the 3-brane has spatial curvature.

In this paper, 
we would like to discuss the self-tuning of general
brane junctions~\cite{Csaki},
a simple brane intersection, and a brane network~\cite{KJLM} 
on the 6-dimensional space-time by introducing a bulk scalar without
bulk potential. 
For the general brane junction models,
 there may exist 
 one fine-tuning among the total brane tensions~\footnote{Although
it is not essentially self-tuning if there was one fine-tuning among
the brane tensions, we still
 call it ``self-tuning'', which 
means the bulk potential for $\phi$ is zero.}. 
However, in the brane intersection models or brane networks, 
where the branes are co-dimension
one hypersurfaces, the brane tensions can be set arbitrarily because the
constraint in the brane junctions is satisfied automatically.
Similar to the 5-dimensional brane models, there exists the
singularity for the metric and bulk scalar
 in the self-tuning of brane networks. And if we want to
regularize the singularity, for example, we require the extra dimensions
be compact and introduce the cut-off branes, we will have the fine-tuning 
among the brane tensions.

In addition,
as we know, our universe may have a very small positive cosmological
constant, so, we discuss the brane network with
de Sitter brane intersections. 
Suppose we have $n$ space-like extra dimensions whose coordinates are
$y^i$ where i=1, 2, ..., $n$, and the branes are co-dimension
one hypersurfaces which are determined by the algebraic equations
$y^i= r$ where r is a real number. If one assumed the metric
\begin{eqnarray}
ds^2&=&\Omega^{-2}\, (- dt^2 + \sum_{i=1}^3 e^{2Ht} \, dx^i\,dx^i +
\sum_{i=1}^n dy^{i2}) ~,~\, 
\end{eqnarray}
it is not difficult for one to show that there does not exist the solution
for $n > 1$~\footnote{For $n=1$ solution, for
example, see ref.~\cite{TN}.}, i. e., we do not have such kind of brane
networks
with de Sitter brane intersections.
In order to obtain the solutions, we introduce a bulk scalar which has 
bulk potential, in other words, we add one degree of freedom to the
system.
We present a 6-dimensional brane network with three (two) 4-branes whose
extra dimension
coordiantes are $y$ and $z$, one
along the y direction and two (one) along the z direction. The solution
has no
singularity and all the brane tensions have similar forms in terms of
the scalar. Similarly, one can also discuss the general brane networks
with
de Sitter or
 Anti-de Sitter brane intersections.

\section{Self-Tuning of the Brane Junctions and Networks}
First, let us discuss the self-tuning of the general brane junctions.
The set-up is given in Fig. 1, we use the metric with 
signature $(-,+,+,+,+,+)$.
 The $6$-dimensional gravitational action describing the system is
\begin{eqnarray}
S = \frac{1}{2\,\kappa^2} \int d^4x dy dz \sqrt{-g} ( R -
 \partial_A\phi \, \partial^A\phi)  
  - \sum_{i=1}^k \int d^4x dy dz \sqrt{-g^{(i)}}  V_i(\phi)
   \delta(\vec{n}_i \cdot \vec{w}) 
  \theta(\vec{l}_i \cdot \vec{w})~, 
\end{eqnarray}
where $\kappa^2= M_X^{-4}$ is the $6$-dimensional coupling constant of 
gravity, $M_X$ is the 6-dimensional Planck scale, and 
$R$ is the curvature scalar. The vectors
\begin{eqnarray}
\vec{n}_i = (-\sin\alpha_i,\cos\alpha_i)\,, \ \ \ 
  \vec{l}_i = (\cos\alpha_i,\sin\alpha_i)\, \ \ \ 
{\rm{and}}\ \ \   \vec{w} = (y,z)
\end{eqnarray}
are defined so that $\delta(\vec{n}_i \cdot \vec{w})$ is the line in the
$y-z$
plane which contains the $i^{th}$ brane and $\theta(\vec{l}_i \cdot
\vec{w})$
serves to cut the irrelevant half of the line. Thus, 
$\delta(\vec{n}_i \cdot \vec{w})\,\theta(\vec{l}_i \cdot \vec{w})$ defines
the 
location of the  $i^{th}$ semi-infinite brane. The
 $5$-dimensional metric on the $i^{th}$ half-brane is
\begin{eqnarray}
g_{AB}^{(i)} \equiv g_{AB}(z=y\,{\rm{tan}}\,\alpha_i)\,.
\end{eqnarray}
Assuming the metric to be
conformally flat, it can be written as
\begin{eqnarray}
ds^2&=&\Omega^{-2}\, (\eta_{\mu\nu}\,dx^{\mu}\,dx^{\nu}+dy^2+dz^2) ~, 
\end{eqnarray}
where  $\Omega \equiv \Omega(y,z)$. Then, the Einstein equations are
\begin{eqnarray}
G_{AB}= \kappa^2\,T_{AB} &=& \partial_A\phi\, \partial_B\phi 
  - \frac{1}{2} \, g_{AB} \,(\partial\phi)^2 
  - \kappa^2\, \Omega^{-1} \, \sum_{i=1}^k \Gamma_{AB}^{(i)} \, V_i(\phi)
\,
  \delta(\vec{n}_i \cdot \vec{w}) \, 
  \theta(\vec{l}_i \cdot \vec{w}) ~,
\end{eqnarray}
where
\begin{eqnarray}
\Gamma_{AB}^{(i)} &=& \pmatrix{-1&&&&&\cr &1&&&&\cr &&1&&\cr &&&1&\cr
                  &&&&\cos^2\alpha_i&\sin\alpha_i\,\cos\alpha_i\cr
                  &&&&\sin\alpha_i\,\cos\alpha_i& \sin^2\alpha_i\cr}\,. 
\end{eqnarray}
These equations can be put in a form amenable to easy solution by
transforming to the conformally related space-time,
\begin{eqnarray}
\tilde{g}_{AB} &=& \Omega^2 \, g_{AB}\,.
\end{eqnarray}

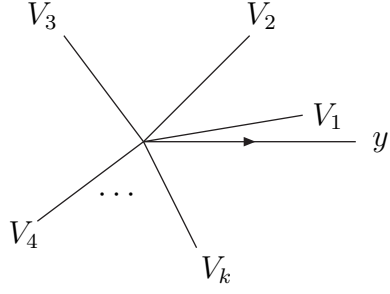
\begin{figure}[h]
\begin{center}
\begin{picture}(200,140)(0,20)
\ArrowLine(100,80)(180,80)
\Line(100,80)(160,90)
\Line(100,80)(140,120)
\Line(100,80)(70,120)
\Line(100,80)(60,50)
\Line(100,80)(120,40)
\Text(190,80)[]{$y$}
\Text(90,60)[]{$\cdots$}
\Text(170,90)[]{$V_1$}
\Text(145,128)[]{$V_2$}
\Text(62,128)[]{$V_3$}
\Text(55,45)[]{$V_4$}
\Text(128,32)[]{$V_k$}
\end{picture}\\
\end{center}
\caption[]{A junction ($3$-brane) of $k$ semi-infinite 
4-branes. The angular positions of 
 branes are measured from the y-axis and the brane tensions are
denoted by $V_i$.}
\label{setup}
\end{figure} 

In six dimensions the Einstein tensor in the two metrics are related by 
\begin{eqnarray}
G_{AB} &=& \tilde{G}_{AB} + 4 \, \Bigl[\Omega^{-1} \,
  \tilde{\nabla}_{A} \, \tilde{\nabla}_{B} \, \Omega
  +\tilde{g}_{AB}\,(- \Omega^{-1} \, \tilde{\nabla}^2 \,\Omega
  + \frac{5}{2}\,\Omega^{-2}\,(\tilde{\nabla}\,\Omega)^2)\Bigr]\,,
\end{eqnarray}
where the covariant derivatives  $\tilde{\nabla}$ are evaluated with
respect to 
the metric $\tilde g$. Since the metric is conformally flat, 
 the covariant derivatives are identical to ordinary derivatives and 
$\tilde G_{AB}=0$. Using above form of the Einstein tensor, 
the Einstein equations are
\begin{eqnarray}
4 \, \frac{\partial^2\Omegain}{\partial y^2} &=& 
  \Omega \Bigl(\dpdy\Bigr)^2 
  +\kappa^2 \sum_{i=1}^k V_i(\phi) \, \sin^2\alpha_i \, \delndx \, \theldx
~, 
\label{dd1}\\
4 \, \frac{\partial^2\Omegain}{\partial z^2} &=&
  \Omega \Bigl(\dpdz\Bigr)^2 
  + \kappa^2 \sum_{i=1}^k V_i(\phi) \, \cos^2\alpha_i \, \delndx \,
\theldx ~, \label{dd2}\\
20 \Bigl(\frac{\partial\Omegain}{\partial y}\Bigr)^2 
  &+& 20 \Bigl(\frac{\partial\Omegain}{\partial z}\Bigr)^2 =
  \Omegainn \, \Bigl(\dpdy\Bigr)^2 
  + \Omegainn \, \Bigl(\dpdz\Bigr)^2 \ \ \ \  \label{dd3}\\
4 \, \frac{\,\partial^2 \Omegain}{\partial y \,\partial z} &=&
  \Omegain\, \dpdy \, \dpdz - \kappa^2\,
  \sum_{i=1}^k \sin\alpha_i\,\cos\alpha_i \, V_i(\phi)\,\delndx \,\theldx
~,
\label{dd4}
\end{eqnarray} 
which must be supplemented with the equation of motion of the scalar 
field $\phi$, 
\begin{eqnarray}
\Omega \, \frac{\partial^2 \phi}{\partial y^2} &
  + & \Omega \, \frac{\partial^2 \phi}{\partial z^2}
  - 4 \, \frac{\partial \Omegain}{\partial y} \, \dpdy
  - 4 \, \frac{\partial \Omegain}{\partial z} \, \dpdz
  = \kappa^2\,\sum_{i=1}^k  \frac{\partial V_i}{\partial \phi}
  \, \delndx \, \theldx ~.
\label{dd5}
\end{eqnarray}
The solution to Eq.s~($\ref{dd1}-\ref{dd5}$) is 
\begin{eqnarray}
\Omegain &=& \Bigl\{ \sum_{i=1}^k (\vec{r}_i \cdot \vec{w}) \,
\theta(\vec{n}_i \cdot \vec{w})
  \, \theta(-\vec{n}_{i+1} \cdot \vec{w}) + C \Bigr\}^{-\frac{1}{4}} ~, \\
\phi &=& -2\,\sqrt{5} \, \ln \Omega\,,
\end{eqnarray}
where $\vec{n}_{k+1} \equiv \vec{n}_{1}$ and $\vec{r}_i=(p_i,q_i)$ are the
integration constants which are not all independent of each other. The 
brane tensions are 
\begin{eqnarray}
\kappa^2\,V_i(\phi) &=& \frac{p_{i}-p_{i-1}}{\sin\alpha_i} \, 
  e^{-\frac{\sqrt{5}}{2} \, \phi} \quad {\rm for~} \sin\alpha_i \ne 0
~,~\,
\end{eqnarray}
or
\begin{eqnarray}
\kappa^2\,V_i(\phi) &=&
  \frac{q_{i-1}-q_{i}}{\cos\alpha_i} \, 
  e^{-\frac{\sqrt{5}}{2} \, \phi} \quad
  {\rm for~} \cos\alpha_i \ne 0 ~.~\,
\end{eqnarray}
Of course, $\sin\alpha_i = 0 $ and $\cos\alpha_i = 0$
imply $ p_{i} = p_{i-1}$ and $ q_{i} = q_{i-1} $, respectively. We obtain 
that there may exist one fine-tuning among
the total brane tensions from Eq.s (17) and (18), which will be
automatically 
satisfied when we consider the
self-tuning of brane intersections or brane networks.
\begin{figure}[t]
\begin{center}
\begin{picture}(200,140)(0,30)

\Line(40,80)(160,80)
\Line(50,40)(150,120)
\Text(170,80)[]{$V_1$}
\Text(158,128)[]{$V_2$}
\Text(130,90)[]{$\alpha$}
\end{picture}\\
\end{center}
\caption[]{A simple brane intersection formed by two infinite 4-branes
intersecting at an
angle $\alpha$.} 
\label{config}
\end{figure}
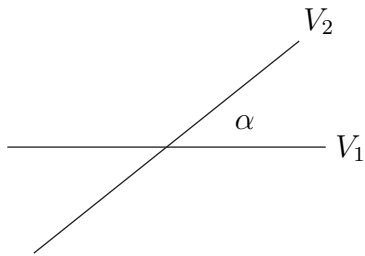

Second, we present a simple brane intersection model which is
an special case of above solution. Suppose
that we have two 4-branes, which are determined by the equations
$y=0$ and $y \sin\alpha = z\cos\alpha$, respectively. The set up is given
at Fig. 2.
The action and Einstein equation can be obtained from above general
discussions,
so, we will not repeat them here. We just give the solution where the
conformal factor
and $\phi$ are
\begin{eqnarray}
\Omegain &=& \big(  a \, | \,z\, |+b \, | \,y\sin\alpha - z\cos\alpha \,| 
   +c \, y+ d \, z + e \big)^{-\frac{1}{4}} ~,~\,
\end{eqnarray}
\begin{eqnarray}
\phi &=& -2\,\sqrt{5} \, \ln \big(  a \, | \,z\, |+b \,
 | \,y\sin\alpha - z\cos\alpha \,| 
   +c \, y+ d \, z + e \big)^{-\frac{1}{4}} ~,~\,
\end{eqnarray}
and the brane tensions are
\begin{equation}
\kappa^2\,V_1(\phi) = - 2\, a \, e^{-\frac{\sqrt{5}}{2} \, \phi} 
~,~ \kappa^2\,V_2(\phi) = - 2\, b \, e^{-\frac{\sqrt{5}}{2} \, \phi}
~.~\,
\end{equation}
We also assume $e~ >~ 0$.
If $a~<~0$ and $b~<~0$, then, two branes have positive tensions and
the brane tensions can be set arbitrarily. However,
$\phi$ will have singularity on some lines and the
conformal factor $\Omega^{-1}$ will vanish there. The singular points form
co-dimension one curves, and can be calculated easily. For example,
assuming $c=d=0$,
there are 4 singular points along the two branes: $(y=-e/a, z=0)$,
$(y=e/a, z=0)$, $(y=-e/b ~\cos\alpha, z=-e/b ~\sin\alpha)$,
and $(y=e/b ~\cos\alpha, z=e/b ~\sin\alpha)$, we can draw 4 straight lines
from  $(y=-e/a, z=0)$ and
$(y=e/a, z=0)$ to $(y=-e/b ~\cos\alpha, z=-e/b ~\sin\alpha)$
and $(y=e/b ~\cos\alpha, z=e/b ~\sin\alpha)$. On those 4 lines, 
$\phi$ is singular and $\Omega^{-1}$ is zero.

Third, we discuss a brane network with four 4-branes on the extra space
manifold
$R^1/Z_2 \times R^1/Z_2$. Two branes along $y$ direction are located
at $y=0$ and $y=y_1$, respectively, and two branes along the $z$ 
direction are located at $z=0$ and $z=z_1$, respectively. The set-up is
given
in Fig. 3. The action for this model is
\begin{eqnarray}
S= S_{\rm Bulk} + S_{\rm Branes} ~,~ \,
\end{eqnarray}
where
\begin{eqnarray}
S_{\rm Bulk} =  {1\over {2\kappa^2}} \int d^4x dy dz \sqrt{-g} ( R -
 \partial_A\phi \, \partial^A\phi ) ~,~\,
\end{eqnarray}
\begin{eqnarray}
S_{\rm Brane} &=& -\int d^4x dy dz  (\sqrt{-g^{(1)}} V_1(\phi) \delta (y)
+ \sqrt{-g^{(3)}} V_3(\phi) \delta (y-y_1))
\nonumber\\&&
-\int d^4x dy dz ( \sqrt{-g^{(2)}} V_2(\phi) \delta (z)
+\sqrt{-g^{(4)}} V_4(\phi) \delta (z-z_1)) ~,~\,
\end{eqnarray}
where $g^{(i)}$ for $i=1, 2, 3, 4$ is the metric on the $i$-th brane,
which
can be obtained by restriction. The detail calculation is similar, so, we 
just give the result. Assuming the confomal metric
\begin{eqnarray}
ds^2&=&\Omega^{-2}\, (\eta_{\mu\nu}\,dx^{\mu}\,dx^{\nu}+dy^2+dz^2) ~,~\, 
\end{eqnarray}
we obtain the solution for conformal factor and $\phi$
\begin{eqnarray}
\Omegain &=& \big(  a \, | \,y-y_1\, |+b \, | \,z-z_1 \,| 
   +c \, y+ d \, z + e \big)^{-\frac{1}{4}} ~,~\,
\end{eqnarray}
\begin{eqnarray}
\phi &=& -2\,\sqrt{5} \, \ln \big(  a \, | \,y-y_1\, |+b \, | \,z-z_1 \,| 
   +c \, y+ d \, z + e \big)^{-\frac{1}{4}} ~,~\,
\end{eqnarray}
and the brane tensions are
\begin{equation}
\kappa^2\,V_1(\phi) = - \, (c-a) \, e^{-\frac{\sqrt{5}}{2} \, \phi}
~,~ \kappa^2\,V_2(\phi) = - \, (d-b) \, e^{-\frac{\sqrt{5}}{2} \, \phi}
~,~\,
\end{equation}
\begin{equation}
\kappa^2\,V_3(\phi) = - 2\, a \, e^{-\frac{\sqrt{5}}{2} \, \phi} 
~,~\kappa^2\,V_4(\phi) = - 2\, b \, e^{-\frac{\sqrt{5}}{2} \, \phi} 
~.~\,
\end{equation}
So, if $c ~>~ a$, $d ~>~ b$ and $e~>~0$, the metric and $\phi$ does not
have
singularity or vanish 
at finite distance from origin, but they are divergent at infinity.
And the brane tensions can be set arbitrarily, for instance,
 if $a~<~0$ and $b~<~0$, the brane tensions $V_3$ and $V_4$ are
positive, and the brane tensions $V_1$ and $V_2$ are negative. In order to
avoid the divergence in the metric and $\phi$, we can introduce two
cut-off
4-branes: $V_5$ which is located at $y=y_2$, and $V_6$ which is located at
$z=z_2$, where $y_2~ > ~y_1$ and $z_2 ~>~ z_1$. So, the extra space
manifold is
$S^1/Z_2\times S^1/Z_2$. Because the tensions for the cut-off 4-branes are
\begin{equation}
\kappa^2\,V_5(\phi) =  \, (c+a) \, e^{-\frac{\sqrt{5}}{2} \, \phi}
~,~ \kappa^2\,V_6(\phi) =  \, (d+b) \, e^{-\frac{\sqrt{5}}{2} \, \phi}
~,~\,
\end{equation}
we will have the fine-tuning among the brane tensions, which is similar to
the 5-dimensional self-tuning models~\cite{FLLN}.

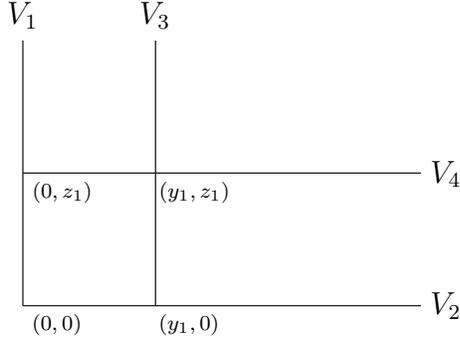
\begin{figure}[t]
%
\begin{center}
\begin{picture}(160,120)(0,0)
\Line(10,10)(160,10)
\Line(10,60)(160,60)
\Line(10,10)(10,110)
\Line(60,10)(60,110)
\Text(10,120)[]{$V_1$}
\Text(60,120)[]{$V_3$}
\Text(170,10)[]{$V_2$}
\Text(170,60)[]{$V_4$}
\Text(23,3)[]{\scriptsize{${(0,0)}$}}
\Text(73,3)[]{\scriptsize{$(y_1,0)$}}
\Text(25,53)[]{\scriptsize{$(0,z_1)$}}
\Text(75,53)[]{\scriptsize{$(y_1,z_1)$}}
\end{picture}
\end{center}
\caption[]{A network of four 4-branes on the 6-dimensional space-time
$M^4\times R^1/Z_2 \times R^1/Z_2$.}
\label{config}
\end{figure}

\section{Brane Network with de-Sitter Brane Intersection}
Because the 4-dimensional cosmological constant we observe is positive
although it is very small, we would like to discuss the brane network with
de Sitter brane intersections.
In order to have the solutions, we introduce one bulk scalar $\phi$ whose
bulk potential does not vanish.
Assume we have three 4-branes, one along the $y$ direction at $y=0$,
two along the $z$ direction at $z=0$ and $z=z_1$. The set-up is given
at Fig. 4 (b). The metrics on the branes can be obtained by restriction
\begin{eqnarray}
g_{AB}^{(1)} \equiv g_{AB}(y = 0), {\rm where} A,B \neq y \, , \\
g_{AB}^{(2)} \equiv g_{AB}(z = 0), {\rm where} A,B \neq z \, , \\
g_{AB}^{(3)} \equiv g_{AB}(z = z_1), {\rm where} A,B \neq z \, .
\end{eqnarray}
And the action for this system is
\begin{eqnarray}
S= S_{\rm Bulk} + S_{\rm Branes} ~,~ \,
\end{eqnarray}
where
\begin{eqnarray}
S_{\rm Bulk} =  \int d^4x dy dz \sqrt{-g} ( {1\over 2} R -
{1\over 2} \partial_A\phi \, \partial^A\phi - \Lambda(\phi) ) ~,~\,
\end{eqnarray}
\begin{eqnarray}
S_{\rm Brane} &=& -\int d^4x dy dz  \sqrt{-g^{(1)}} V_1(\phi) \delta (y)
\nonumber\\&&
-\int d^4x dy dz (\sqrt{-g^{(2)}} V_2(\phi) \delta (z)
+\sqrt{-g^{(3)}} V_3(\phi) \delta (z-z_1)) ~.~\,
\end{eqnarray}

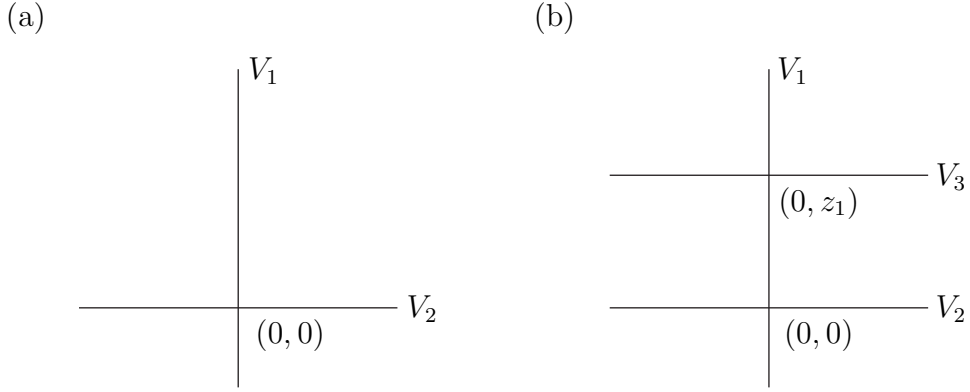
\begin{figure}[t]
\begin{center}
\begin{picture}(400,140)(0,10)
\Line(40,50)(160,50)
\Line(100,20)(100,140)
\Text(170,50)[]{$V_2$}
\Text(110,140)[]{$V_1$}
\Text(120,40)[]{$(0,0)$}
\Text(20,160)[]{(a)}
\Line(240,50)(360,50)
\Line(240,100)(360,100)
\Line(300,20)(300,140)
\Text(310,140)[]{$V_1$}
\Text(370,50)[]{$V_2$}
\Text(370,100)[]{$V_3$}
\Text(320,40)[]{$(0,0)$}
\Text(320,90)[]{$(0,z_1)$}
\Text(220,160)[]{(b)}
\end{picture}\\
\end{center}
\caption[]{(a) Two 4-branes with de Sitter brane intersection;
  (b) Three 4-branes with two de Sitter brane intersections.}
\end{figure}

With the following conformal metric 
\begin{eqnarray}
ds^2&=&\Omega^{-2}\, (- dt^2 + \sum_{i=1}^3 e^{2Ht} \, dx^i\,dx^i +
dy^2+dz^2) ~, 
\end{eqnarray}
we obtain the Einstein equations
\begin{eqnarray}
4 \, \frac{\partial^2\Omegain}{\partial y^2} &=& 
  \Omega \Bigl(\dpdy\Bigr)^2 + V_1(\phi) \, \delta(y) + 
  3 H^2 \, \Omega~, 
  \label{aa1}\\
4 \, \frac{\partial^2\Omegain}{\partial z^2} &=&
  \Omega \Bigl(\dpdz\Bigr)^2 + V_2(\phi) \, \delta(z) +
V_3(\phi) \, \delta(z-z1)+
  3 H^2 \, \Omega~, 
  \label{aa2}\\
20 \Bigl(\frac{\partial\Omegain}{\partial y}\Bigr)^2 
  &+& 20 \Bigl(\frac{\partial\Omegain}{\partial z}\Bigr)^2 =
  \Omegainn \, \Bigl(\dpdy\Bigr)^2 
  + \Omegainn \, \Bigl(\dpdz\Bigr)^2 
  - 2 \Lambda(\phi) + 18H^2\,\Omegainn ~,
  \label{aa3}\\
4 \, \frac{\,\partial^2 \Omegain}{\partial y \,\partial z} &=&
  \Omegain\, \dpdy \, \dpdz  ~,  
  \ \ \ \  
  \label{aa4} \\
\end{eqnarray}
and the equation of motion for $\phi$
\begin{eqnarray}
\Omegainn \, \frac{\partial^2 \phi}{\partial z^2}
  + \Omegainn \, \frac{\partial^2 \phi}{\partial z^2}
  &=& 
  + 4 \, \Omega \,\frac{\partial\Omega}{\partial y} \,
  \frac{\partial \phi}{\partial y}
  + 4 \, \Omega \, \frac{\partial\Omega}{\partial z}\,
  \frac{\partial \phi}{\partial z}  
  + \Omega \, \frac{\partial V_1(\phi)}{\partial \phi} \, \delta(y)
\nonumber\\&&
  + \Omega \, \frac{\partial V_2(\phi)}{\partial \phi} \, \delta(z)
  + \Omega \, \frac{\partial V_3(\phi)}{\partial \phi} \, \delta(z-z_1)
  + \frac{\partial \Lambda(\phi)}{\partial \phi} ~.~\,
  \label{aa5}
\end{eqnarray} 

If $V_3(\phi) =0$, i. e., there are two 4-branes and the set-up is given
in
Fig. 4 (a). The conformal factor and $\phi$ are
\begin{eqnarray}
\Omega &=& \exp \{\,\frac{3}{8}\,H^2\,[\,(\,|\,y\,| + c_1)^2 + 
  (\,|\,z\,| + c_2)^2 + c_3\,]\,\} ~,~\,
\end{eqnarray}
\begin{eqnarray}
\phi &=& \frac{3}{4}\,H^2\,[\,(\,|\,y\,| + c_1)^2 + 
  (\,|\,z\,| + c_2)^2 + c_3\,]~,~\,
\end{eqnarray}
the bulk potential for $\phi$ is
\begin{eqnarray}
\Lambda(\phi) &=& 9 (1+{1\over 2} H^2 c_3) H^2\, e^{\phi}\, 
- 6  H^2\, \phi \, e^{\phi} ~,~\,
\end{eqnarray}
and the brane tensions are
\begin{equation}
V_1(\phi)~=~6~e^{\phi/2}
~,~ V_2(\phi)~=~6~e^{\phi/2}
~.~\,
\end{equation}
So, both branes have positive tensions.

Now, we consider the case $V_3(\phi) \ne 0$, because
we need at least three 4-branes to solve the gauge hierarchy problem.
The conformal factor and
$\phi$ are
\begin{eqnarray}
\Omega &=& \exp \{\,\frac{3}{8}\,H^2\,[\,(\,|\,y\,| + c_1)^2 + 
  (\,|\,z\,| - |\,z-z_1\,| - z + c_2)^2 + c_3\,]\,\} ~,~\,
\end{eqnarray}
\begin{eqnarray}
\phi &=& \frac{3}{4}\,H^2\,[\,(\,|\,y\,| + c_1)^2 + 
  (\,|\,z\,| -|\,z-z_1\,| - z + c_2)^2 + c_3\,]~,~\,
\end{eqnarray}
the bulk potential for $\phi$ is
\begin{eqnarray}
\Lambda(\phi) &=& 9 (1+{1\over 2} H^2 c_3) H^2\, e^{\phi}\, 
- 6  H^2\, \phi \, e^{\phi} ~,~\,
\end{eqnarray}
and the brane tensions are
\begin{equation}
V_1(\phi)~=~6~H^2~e^{\phi/2}
~,~ V_2(\phi)~=~6~H^2~e^{\phi/2}
~,~\,
\end{equation}
\begin{eqnarray}
V_3(\phi) ~=~ - ~6~H^2~e^{\phi/2} ~.~\,
\end{eqnarray}
Thus, the third brane has negative tension. By the way, 
all the brane tensions have similar forms in terms of
the scalar.

Similarly, we can discuss the general brane networks with de Sitter or
 Anti-de Sitter brane intersections.

\section{Conclusion}
We study the self-tuning of general brane junctions and brane
networks on the 6-dimensional space-time. For the general brane junctions,
 there may exist one fine-tuning among the brane tensions. 
For the brane networks,
similar to the 5-dimensional self-tuning brane models, the brane
tensions can be set arbitrarily and 
there exists the singularity for the metric and bulk scalar. And if we
want to
regularize the singularity, we will introduce the fine-tuning 
among the brane tensions. In addition, because the 4-dimensional
cosmological
constant we observe may be positive and very small,
we discuss the brane network
with de Sitter brane intersections by introducing a bulk scalar.

\section*{Acknowledgments.}
We would like to thank D. Mafatia for collaboration in the early stage of
this project.
The work of J. J.  was supported in part by DOE grant No. 
DE-FG02-95ER40896 and in part by the Wisconsin Alumni Research Foundation.
And
the work of T. L. was supported by DOE grant No. DOE-EY-76-02-3071.  
\\
\\

\end{document}